\documentclass[letterpaper]{article} 
\usepackage[preprint]{aaai2027} 
\usepackage[hyphens]{url}  
\usepackage{graphicx} 
\usepackage{array}  
\usepackage{natbib}  
\usepackage{caption} 
\usepackage{algorithm}
\usepackage{algorithmic}

\usepackage{newfloat}
\usepackage{listings}
\DeclareCaptionStyle{ruled}{labelfont=normalfont,labelsep=colon,strut=off} 
\floatstyle{ruled}
\newfloat{listing}{tb}{lst}{}
\floatname{listing}{Listing}

\usepackage{xcolor}
\usepackage[useregional]{datetime2}
\usepackage{algorithmic}
\usepackage{textcomp}
\usepackage{xcolor}
\usepackage{amsmath}
\usepackage{array}
\usepackage{tabularray}
\usepackage{booktabs}
\usepackage{multirow}
\usepackage{listings}
\usepackage{subcaption}
\usepackage{enumitem}
\usepackage{pifont}
\usepackage{booktabs}
\usepackage{multirow}
\usepackage{xcolor}
\usepackage{colortbl}
\usepackage{siunitx}   
\usepackage{amssymb}
\usepackage[export]{adjustbox}
\usepackage{tikz}
\usetikzlibrary{shapes.geometric, arrows.meta, positioning}
           
\newcommand{\wrt}{\mbox{{w.r.t}}}

\newcommand{\prompt}{\mbox{{\scshape 
NLS}}}

\newcommand{\promptzero}{\mbox{{\scshape NLS@0}}}

\newcommand{\icl}{\mbox{{\scshape NLSC}}}

\newcommand{\iclzero}{\mbox{{\scshape NLSC@0}}}

\newcommand{\iclfive}{\mbox{{\scshape NLSC@5}}}

\newcommand{\baseline}{\mbox{{\scshape NL}}}

\newcommand{\pname}{\mbox{{\scshape 
Laude}}}

\newcommand{\buggycode}{\mbox{{$\mathcal{D}_\mathcal{B}$}}}

\newcommand{\correctcode}{\mbox{{$\mathcal{D}_\mathcal{C}$}}}

\newcommand{\unittest}{\mbox{{$\mathcal{U}$}}}

\newcommand{\failtrace}{\mbox{{$\mathcal{T}_f$}}}

\newcommand{\passtrace}{\mbox{{$\mathcal{T}_p$}}}

\newcommand{\trace}{\mbox{{$\mathcal{T}$}}}

\newcommand{\oracle}{\mbox{{$\mathbb{O}$}}}

\newcommand{\curly}{\mathrel{\leadsto}}

\newcommand{\signature}{\mbox{{$\mathcal{S}_\mathcal{D}$}}}

\newcommand{\generator}{\mbox{{$\mathbf{G}_\Theta$}}}

\newcommand{\bem}[1]{{\bf\em #1}}

\newcommand{\eg}{\mbox{{\em e.g.}}}
\newcommand{\ie}{\mbox{{\em i.e.}}}
\newcommand{\cf}{\mbox{{c.f.}}}

\newcolumntype{C}[1]{>{\centering\arraybackslash}p{#1}}
\newcolumntype{L}[1]{>{\arraybackslash}p{#1}}

\usepackage{listings}
\definecolor{dmlgreen}    {RGB}{51,  160,  44}
\definecolor{dmlblue}     {RGB}{31,  120, 180}
\definecolor{dmlred}      {RGB}{202,   0,  32}

\definecolor{codegreen}{rgb}{0,0.6,0}
\definecolor{codegray}{rgb}{0.5,0.5,0.5}
\definecolor{codepurple}{rgb}{0.58,0,0.82}
\definecolor{backcolour}{rgb}{0.95,0.95,0.92}

\lstdefinestyle{mystyle}{
    backgroundcolor=\color{backcolour},   
    commentstyle=\color{codegreen},
    keywordstyle=\color{blue},
    numberstyle=\tiny\color{codegray},
    stringstyle=\color{codepurple},
    basicstyle=\ttfamily\footnotesize,
    breaklines=true,                 
    captionpos=b,                    
    numbers=left,                    
    numbersep=5pt,                  
    showspaces=false,                
    showstringspaces=false,
    showtabs=false,                  
}

\lstdefinestyle{customtext}
{
  backgroundcolor=\color{backcolour}, 
  basicstyle=\ttfamily\footnotesize,
  commentstyle=\color{codegreen},
  keywordstyle=\color{magenta}\bfseries, 
  stringstyle=\color{codepurple},  
  numberstyle=\tiny\color{codegray},
  breaklines=true,
  captionpos=b,
  keepspaces=true,
  numbers=left,
  numbersep=5pt,
  frame=single,                    
  rulecolor=\color{codegray},      
  showstringspaces=false,
  tabsize=2,
  xleftmargin=0.5em,               
  framexleftmargin=1em,
  moredelim=[s][\color{codepurple}]{<}{>},
  moredelim=[s][\color{red!50!black}]{[}{]},
  keywords={Module:, Input, Output, Verilog, Task:, Requirements:, Output, Generate, Task, Module, Current, Buggy, Code, Under, Analysis, Problem, Simulation, Artifacts, Debugging, Strategies, Results},
  keywordstyle=\bfseries\color{blue!40!black},
  columns={fullflexible},
}%

\usepackage{booktabs}
\usepackage{multicol}
\usepackage{multirow}

\title{$\pname$: \underline{L}LM-\underline{A}ssisted \underline{U}nit Test Generation and \underline{D}ebugging of Hardware D\underline{e}signs}
\title{$\pname$: \underline{L}LM-\underline{A}ssisted \underline{U}nit Test Generation and \underline{D}ebugging of Hardware D\underline{e}signs}
\author {
    Deeksha Nandal\textsuperscript{\rm 1}\equalcontrib,
    Riccardo Revalor\textsuperscript{\rm 1}\equalcontrib,
    Soham Dan\textsuperscript{\rm 2}\corresponding,
    Debjit Pal\textsuperscript{\rm 1}\corresponding
}
\affiliations {
    \textsuperscript{\rm 1}University of Illinois Chicago\\
    \textsuperscript{\rm 2}Scale AI\\
    dnanda6@uic.edu, rreva@uic.edu, sdan021@gmail.com, dpal2@uic.edu
}
\begin{document}

\maketitle

\begin{abstract}


Unit tests are critical in the hardware design lifecycle to ensure that component design modules are functionally correct and conform to the specification before being integrated at the system level. Thus, developing unit tests 
targeting various design features requires a deep understanding of the design 
functionality and creativity. When one or more unit tests expose a design failure, the debugging engineer must diagnose, 
localize, 
and debug the failure to ensure correctness of the design, which is often an intense process. In this work, we introduce $\pname$, 
a unified unit-test generation and debugging framework for hardware designs that cross-pollinates the semantic understanding of the design source code with the Chain-of-Thought (CoT) reasoning capabilities of foundational Large-Language Models (LLMs). $\pname$ integrates prompt engineering and design execution information to enhance its unit test generation accuracy and code debuggability. 
We apply $\pname$  with closed 
and open-source LLMs to a large corpus of buggy hardware design codes derived from the VerilogEval dataset, where generated unit tests detected 
bugs in up to 100\% and 93\% of combinational and sequential designs and 
debugged up to 93\% and 84\% 
of combinational and sequential designs, respectively. 
Beyond these injected bugs, $\pname$'s simulation-grounded unit tests perform better than the LLM4DV stimulus generator across all designs and on NVIDIA's CVDP benchmark.
This establishes that $\pname$ generalizes from synthetic to real, naturally occurring hardware bugs.

\end{abstract}


\section{Introduction
}\label{sec:intro}


As hardware systems become ubiquitous and more complex, the verification of functional correctness of such systems is becoming the biggest bottleneck ~\cite{veira2018date}. Debugging functional bugs in hardware designs written in hardware description languages (HDLs) is 
an iterative time-consuming process. Verification engineers identify design bugs and refine designs using a test-based environment where a set of valid inputs is provided to the designs and reasoned over failures to address the root cause. Such inputs are called unit tests and often require a deep understanding of design architecture and functionality. Unlike software, HDLs behavior unfolds over clock cycles and depends on registers, finite-state machines (FSMs), and concurrent logic. As a result, creating unit tests that reliably expose functional errors requires significant manual effort and experience.

\begin{figure}
    \begin{minipage}[t]{0.41\columnwidth}
    \begin{subfigure}[t]{\columnwidth}
\begin{lstlisting}[language=Verilog, 
                 %linewidth=2.8in, 
                 framexleftmargin=2pt,
                 framexrightmargin=2pt,
                 basicstyle=\ttfamily\footnotesize,
                 ]
module fa(a, b, 
c, s, 
cout);
input a, b, c;
output s, cout;

s = a ^ b ^ c;
cout = ab + bc + ca;
endmodule
\end{lstlisting}
    \caption{}
    \label{fig:comb_fa}
    \end{subfigure}
    \begin{subfigure}[t]{\columnwidth}
        \includegraphics[scale=0.58,center]{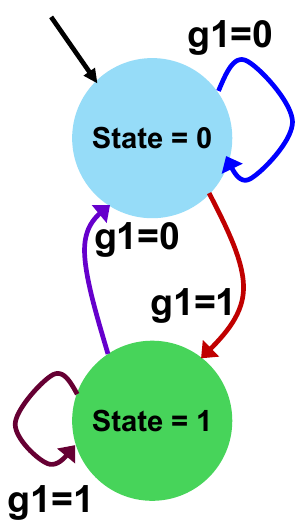}
    \caption{}
    \label{fig:seq_arb_fsm}
    \end{subfigure}
    \end{minipage}
    \hfill
    \begin{minipage}[b]{0.55\columnwidth}
    \begin{subfigure}[t]{\columnwidth}
\begin{lstlisting}[language=Verilog, 
                 %linewidth=2.8in, 
                 framexleftmargin=2pt,
                 framexrightmargin=2pt,
                 basicstyle=\ttfamily\footnotesize,
                 ]
  module arb2(clk, rst, r1, 
  r2, g1, g2);
  input clk, rst, 
  r1, r2;
  output reg g1, g2;
  reg State;
  always @(posedge clk or 
  posedge rst)
    if(rst)
       State <= 0;
    else
       State <= g1;
  always @(*)
    if (State)
    begin
       g1 = r1 & r2;
       g2 = r2;
    end
    else
    begin
      /* buggy code */
      g1 = ~r1 & r2;
      g2 = r1;
    end
  endmodule
\end{lstlisting}
                 \vspace{-3mm}
    \caption{}
    \label{fig:seqc_arb2}
    \end{subfigure}
    \end{minipage}    
    \caption{(\subref{fig:comb_fa}) A 1-bit full adder.~(\subref{fig:seq_arb_fsm}) An FSM capturing correct behavior of a 2-port arbiter.~(\subref{fig:seqc_arb2}) A buggy arbiter.}
    \label{fig:sva_example}
    \vspace{-7mm}
\end{figure}

Consider a 1-bit full adder ({\tt FA}) of Figure~\ref{fig:comb_fa}, the FSM of a 2-port arbiter ({\tt Arb}) of Figure~\ref{fig:seq_arb_fsm}, and a buggy {\tt Arb} of Figure~\ref{fig:seqc_arb2}. For {\tt FA}, outputs ({\tt s, cout}) depend only on inputs ({\tt a, b, c}), much like a software function; for {\tt Arb}, however, outputs ({\tt g1, g2}) also depend on {\tt State}, clock, and an asynchronous reset. Lines 22-23 contain the bug that causes {\tt Arb} to diverge from the desired FSM behavior, but only when {\tt State = 0}: sensitizing it requires a unit test that drives {\tt Arb} to that state, demanding creativity and a deep understanding of the FSM.


Recent research has explored the use of LLMs for various hardware design and verification tasks 
to improve syntactic correctness and functional coherence. It remains underexplored whether LLMs 
can reason and generate test inputs that 
can trigger and propagate the bugs to observable outputs. This is 
challenging since it requires reasoning across clock cycles, 
interactions in FSMs, and is heavily dependent on an implementation (\cf, Figure~\ref{fig:sva_example}). Additionally, a unit test, upon failure, should produce a trace 
that is sufficiently divergent 
and contain a distinct failure signature for successful debugging. However, there is an acute lack of research that explores the reasoning abilities of LLMs to debug design by developing unit tests.

In this work, we propose $\pname$ that combines semantic understanding of design code, advanced reasoning such as CoT, prompt engineering, and simulation feedback to address challenges for unit test generation and source code debugging for hardware designs. To evaluate $\pname$, we use a buggy dataset derived from 
VerilogEval~\cite{liu2023iccad} containing 1,560 buggy codes (10 buggy codes for each of the 156 problems) by randomly injecting one representative bug from a set of commonly occurring bug types. Our experiments show that unit tests detected 
bugs in up to 100\% and 93\% of combinational and sequential designs, and 
debugged up to 93\% and 84\% 
of combinational and sequential designs, respectively, thus establishing the effectiveness of $\pname$. 

\noindent\textbf{Contributions.} We make the following contributions. (i) We propose $\pname$, a unified, closed-loop framework that couples LLM reasoning (CoT, prompt engineering, and in-context learning) with simulation feedback to jointly perform unit test generation and iterative debugging of hardware designs. (ii) On a controlled benchmark of 1,560 buggy designs derived from VerilogEval, $\pname$ generates unit tests that are simultaneously sensitive (high Attack Rate) and specific (high Divergence Rate), and debugs up to 93\% and 84\% of combinational and sequential designs. (iii) Against LLM4DV, the closest prior LLM-based stimulus generator, $\pname$'s simulation-grounded unit tests perform better on five structurally diverse real designs across every metric. (iv) On the CVDP benchmark, we move from synthetic to naturally occurring bugs: since no golden RTL is provided, any LLM-generated RTL that fails at least one test is, by construction, buggy, and $\pname$'s debugging loop repairs these real bugs over successive rounds. This last result directly addresses the concern that bug injection in VerilogEval is synthetic, by demonstrating that $\pname$ operates on bugs that emerge organically from real design generation.

The rest of the paper is organized as follows: 
background and preliminaries, 
methodology,
experimental setup, results,
related work, and conclusion. 
\section{Background and Preliminaries}\label{sec:background}

\noindent \bem{Notations}. We consider SystemVerilog (SV) as the primary HDL; however, our technique can be 
extended to other HDLs, \eg, Verilog, VHDL, etc. We consider a hardware design $\mathcal{D}$ 
in SV 
as an SV program for 
source code analysis. 
Let $\mathcal{V}$ be the set of design variables, $\mathcal{I} \subset \mathcal{V}$ be the set of input variables, and $\mathcal{O} \subset \mathcal{V}$ be the set of output variables. A unit test $\unittest$ for a design $\mathcal{D}$ is defined as the binary 
value ($\mathbb{B} \in [0, 1]$) assignment to all input variables over $\mathcal{N}$ clock cycles, \ie, $\mathcal{I} \mapsto \mathbb{B}^{m \times \mathcal{N}}$ where $|\mathcal{I}| = m$. 
A \bem{simulation run} 
with respect to ($\wrt$) 
$\unittest$ is a time-stamped $\mathcal{N}$ cycle sequence of design variables $v \in \mathcal{V}$ values from input to output. A \bem{simulation trace} $\trace$ $\wrt$ 
$\unittest$ 
is the set of all simulation runs 
for all inputs going to outputs. A simulation run is a \bem{failure run} 
$\wrt$ a variable $v \in \mathcal{O}$ if $\exists n \in \mathcal{N}$ such that $v^n \neq v^n_{expected}$ (\ie, a value mismatch) where $v^n_{expected}$ is the expected value of variable $v$ at cycle $n$ and $v^n \in \failtrace$ (failure trace). We annotate such a design code as a buggy code $\buggycode$. A simulation run is a \bem{passing run} 
if $\forall v \in \mathcal{O}$, $\forall n \in \mathcal{N}, v^n == v^n_{expected}$ (\ie, value match) and produces a passing trace $\passtrace$. We annotate such a design code as the correct code $\correctcode$. 
We assume 
an oracle $\oracle$ 
(\eg, a 
functional 
model of 
design $\mathcal{D}$) that can simulate the correct design behavior. This is a reasonable assumption since such models are usually developed during 
conceptualization. 

\smallskip
\noindent \bem{VerilogEval} is a hardware design benchmarking suite targeted toward evaluating LLM's performance in the context of hardware design and verification~\cite{liu2023iccad}. VerilogEval consists of 156 design problems from a 
Verilog instructional website~\cite{hdlbits2025} 
segregated into a set of 
82 combinational designs 
and 74 
sequential designs with complex 
finite state machines (FSMs). Each design problem consists of a reference SV code implementation and a reference testbench to simulate the correct design behavior. 

{The \bem{Comprehensive Verilog Design Problems} (CVDP) benchmark 
is a framework for evaluating LLMs and agentic solutions on hardware design and verification challenges~\cite{pinckney2025arxiv}, comprising 783 problems across 13 categories spanning code generation and code comprehension, across both open and commercial design settings. Each datapoint provides a hidden test harness that validates the correctness of a generated design, but the golden RTL is not exposed to the solver. CVDP therefore serves as a real-world setting in which bugs arise naturally: any LLM-generated RTL that fails at least one harness test is, by construction, a buggy design.}

{\bem{LLM4DV} is an open-source benchmarking framework that orchestrates LLMs for automated hardware test stimuli generation in the context of design verification (DV)~\cite{zhang2025arxiv}, evaluating six different LLMs across six prompting improvements over eight hardware designs. We use the LLM4DV designs as a 
baseline against $\pname$'s simulation-grounded unit tests.}
\section{$\pname$ Methodology}\label{sec:methodology}

\noindent \bem{Problem Statement.} Given a natural language 
description $\mathcal{L}$ of a hardware design $\mathcal{D}$ and an oracle $\oracle$, we 
generate a unit test $\unittest$ such that when $\unittest$ applied to 
$\mathcal{D}$, it produces an output that is consistent with $\oracle$'s output on $\unittest$ (\ie, $\passtrace$). 
Our problem 
has the following assumptions 
- (i) we consider a {\em parameterized unit test} $\unittest(i_1, i_2, \ldots, i_m)$ 
where we treat each design input $i_k \in \mathcal{I}$ as a 
parameter and assign a set of binary 
values 
to each 
input, and (ii) we establish 
{\em functional correctness} 
of the design by simulating it with a 
set of $\unittest(i_1, i_2, \ldots, i_m)$ and ensuring the output matches the expected output. For brevity, we will denote $\unittest(i_1, i_2, \ldots, i_m)$ as $\unittest$. 

\begin{figure}
     \centering
     \includegraphics[scale=0.32]{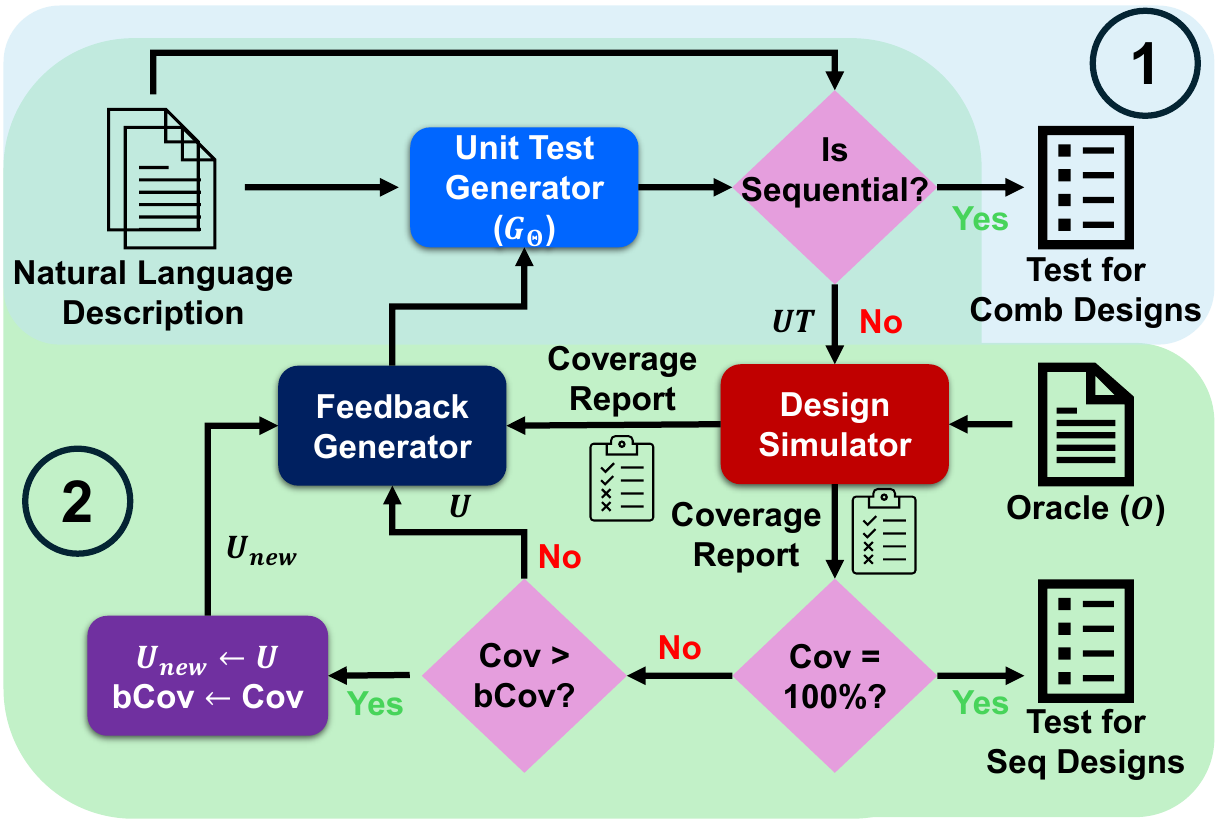}
     \caption{{\bf Iterative unit test generation of $\pname$}. $\generator$: Generator of the unit test $\unittest$. {\bf Cov}: Coverage information. {\bf bCov}: Best coverage until current iteration. $\unittest$: Unit test. {\bf Feedback Generator}: Combines coverage information and the unit test for the next iteration.}
     \label{fig:utgen_flow}
     \vspace{-5mm}
\end{figure}

\subsection{Unit Test Generation Technique}\label{sec:unit_test}

We consider a unit test $\unittest$ to be valid for design $\mathcal{D}$ if $\correctcode(\unittest) \curly \passtrace$.
We plan to develop a unit test generator $\generator$ with hyperparameters $\Theta$ to generate a valid unit test $\unittest$ such that $\generator(\mathcal{L}) \mapsto \unittest$ without any designer intervention.
However, since a hardware design has a specific set of (possibly multi-bit) input and output signals (\eg, {\tt request}, {\tt data}, {\tt acknowledge}), it is important that we provide the design input/output description ({\em signature}) $\signature$ as input to the generator $\generator(\mathcal{L}, \signature)$; such signatures can be easily obtained from the design source code via lightweight source code parsing~\cite{slang2025}.
The key usage of a unit test is to establish the functional correctness of a specific HDL implementation of a hardware design, so we also include the (potentially) buggy implementation $\buggycode$ in the generator input $\generator(\mathcal{L}, \signature, \buggycode)$: by exposing the generator $\generator$ to $\buggycode$, it can semantically analyze the source code and identify a set of binary input values that will produce a failure trace exposing the deviation from the design intent expressed in $\mathcal{L}$, \ie, $\buggycode(\unittest) \curly \failtrace$.
We used various closed- and open-source foundational LLMs for the experiment results.
A key objective of the unit test generation technique is that the unit test should produce a significant divergent trace $\failtrace$ compared to the output of $\oracle$ when applied to the buggy code $\buggycode$.
\bem{In our analysis}, we evaluate and compare $\generator(\mathcal{L}, \signature)$ ($\prompt$) and $\generator(\mathcal{L}, \signature, \buggycode)$ ($\icl$) as generator techniques; our initial experiments showed that $\baseline$ often generates unit tests that are syntactically and/or semantically incorrect and cannot be simulated, so we leave a full comparison with $\baseline$ to future work. Figure \ref{fig:utgen_flow} shows the iterative unit test generation flow, which uses a structured prompt.

For combination circuits (shown in blue with {\large \ding{172}}), generation of $\unittest$ is a one-shot process; for sequential circuits it is iterative (shown in green with {\large \ding{173}}). The $\generator$ aims to generate a unit test that has significant design coverage ($\wrt$ oracle $\oracle$), \eg, FSM coverage, accepting a newly generated unit test only when the coverage increases; we provide the coverage report to $\generator$ to aid in its generation process.

\subsection{Unit Test-Assisted Debugging Technique}\label{sec:debugging}

\begin{figure}
     \centering
     \includegraphics[scale=0.32]{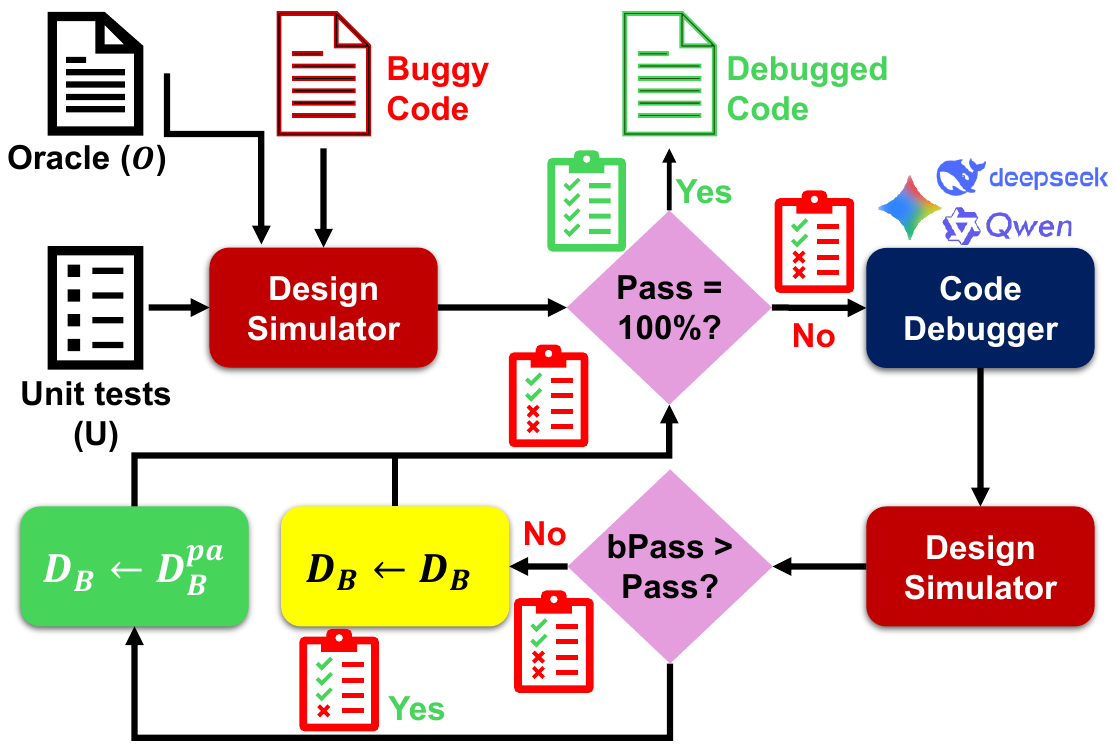}
     \caption{
     {{\bf Unit-test driven iterative debugging of $\pname$}. {$\buggycode$}: Buggy code. {$\mathcal{D}_{\cal B}^{pa}$}: Patched code after debugging with LLM. {\bf Pass}: Fraction of unit tests passed. {\bf bPass}: Best Pass} until current iteration. 
     }
     \label{fig:debug_flow}
     \vspace{-5mm}
\end{figure}

A concern for using the automatically generated unit test $\unittest$ with debugging is that the unit tests can be noisy, \ie, inaccurate in two ways (i) $\unittest$ fails to {\em sensitize} the bug in $\buggycode$, creating a passing trace $\passtrace$ instead of exposing the bug, pointing to {\em insufficient coverage of the design source code}, or (ii) the unit test activates the bug but its effect gets {\em masked} before propagating to an output, indicating the generator's {\em insufficient semantic understanding of the design source code}. Both errors are \bem{detrimental} for debugging, as they could remove the likely buggy code $\buggycode$ from the debugging process early on, without actually fixing the bug.
To tackle these problems, 
we propose 
an {\em iterative} debugging approach as shown in 
Figure ~\ref{fig:debug_flow}.

We begin with the potentially bugged code $\buggycode$, a unit test that has generated a failed trace $\failtrace$, and a summary of the mismatches that are symptomatic of failure. Intuitively, the mismatches provide the failure signature that needs to be investigated to localize the buggy code. 
We use a structured prompt 
(the debugging prompt template is provided in the supplementary material), to incorporate the necessary information for the LLM
and ask it to generate a likely debugged code $\mathcal{D}_{\cal B}^{pa}$. However, $\mathcal{D}_{\cal B}^{pa}$
may introduce additional bugs (since LLMs are probabilistic models)
or may not fix the original bug at all. Hence, instead of accepting the patched
code immediately, we rerun all unit tests on $\mathcal{D}_{\cal B}^{pa}$
and only accept $\mathcal{D}_{\cal B}^{pa}$
if it increases the fraction of
unit tests that produce passing trace $\passtrace$.
The iteration continues until all unit tests pass
or a fixed
number of iterations is completed (we set it to {\bf 5}).

\section{Experimental Setup}\label{sec:exp_setup}


\noindent \bem{Datasets.} We evaluate $\pname$ on three benchmarks of increasing complexity consisting of synthetic and realistic design bugs. We derived VerilogEvalD by systematically injecting one functional bug per design of the VerilogEval dataset from a database of commonly occurring functional bugs. We used a prompting technique to generate bugs from Gemini-2.5 Pro. We instructed the model to generate ten different types of bugs to ensure variety. We ran a simulation for all the faulty codes, and they all passed synthesis. All injected bugs are semantic rather than syntactic. 
Additionally, we have also used LLM4DV and CVDP benchmarks as summarized in Table~\ref{tab:dataset_stats}. 


\begin{table}[t]
\centering
\small
\setlength{\tabcolsep}{1pt}
\begin{tabular}{lccl}
\toprule
\textbf{Dataset} & \textbf{\#Problems} & \textbf{\#Buggy Instances} & \textbf{Bug Source} \\
\midrule
VerilogEvalD 
& 156 
& 10  & Synthetic \\
LLM4DV      & 5                          & 10                        & Synthetic \\
CVDP        & 783 (13 categories)        & 10 & Real-world \\
\bottomrule
\end{tabular}
\caption{Overview of the three benchmarks used to evaluate \pname, spanning synthetic, competitive, and real-world bug sources.}
\label{tab:dataset_stats}
\end{table}

\smallskip
\noindent \bem{LLMs Used.} We have used multiple open-source and closed-source reasoning and coding 
LLMs to evaluate the effectiveness of $\pname$.
DeepSeek-R1-Distill-Qwen-32B \cite{deepseek_ai_2025_deepseek_r1_distill_qwen32b} is fine-tuned from the Qwen-2.5-32B \cite{qwen2025qwen25technicalreport} base model using a dataset generated by the DeepSeek-R1 (671B) model. We 
selected this model to \bem{test} {\em if advanced reasoning 
capabilities are effective for hardware design and validation tasks} in a locally deployable fashion. We included Qwen-2.5 Coder-32B-Instruct~\cite{hui2024qwen2} to \bem{test} {\em how a model specialized for code generation 
would perform against more generalist models.} 
For DeepSeek and Qwen we used the original 
FP16 precisions, and \texttt{vLLM} ~\cite{kwon2023sosp} as the inference engine.


\smallskip 
\noindent \bem{Hardware Platform Used.}  We used 
(i)  
two 
NVIDIA RTX 6000 Ada GPUs, 
leveraging tensor parallelism \cite{ narayanan2021efficient} and (ii) 
a 
NVIDIA H200 GPU, leveraging its massive high-bandwidth memory 
to maximize execution speed.

\smallskip
\noindent \bem{Hyperparameters.}
For all test generator and debugging techniques, we maintained consistent hyperparameters ($\Theta$) to ensure a fair comparison: {\em temperature} was set to 0.8 to {\em balance creativity with determinism}.
The {\em output token generation limits} were set based on the deployment infrastructure: for open-source models we used a dual strategy of 2,048 tokens for $\icl$ (to accommodate code-specific reasoning) and 512 for $\prompt$ and $\baseline$; for closed-source models we used the API's maximum token limit, delegating resource management to the provider's infrastructure.
We restricted the {\em input context window} to $16,384$ tokens for all open-source models well below the theoretical maximum, but sufficient to contain the largest design and testbench codes in our datasets with a reasonable Key-Value (KV) cache footprint and no Out-of-Memory errors. We set the {\em maximum GPU memory} utilization to 0.95 for \texttt{vLLM}, with the remaining 0.05 acting as swap space for temporary tensors and KV pairs.

\smallskip

\noindent \bem{Metrics.} We measure the effectiveness of $\pname$ by quantifying the {\em sensitivity} (\bem{Attack Rate} (AR)) and {\em specificity} (\bem{Divergence Rate} (DR)) of the 
unit tests. A unit test is \bem{sensitive} {\em if} it can detect many different bugs in a design and creates a $\failtrace$. A unit test is \bem{specific} {\em if} it can generate a failure trace $\failtrace$ that is divergent compared to a passing trace $\passtrace$ and carries a unique failure signature for a bug. 
We use  
{\em Divergent Attack} (DA) to quantify the goodness of a unit test $\wrt$ 
sensitivity and specificity. 

\smallskip

\noindent {\bf Attack Rate (AR).} Attack Rate quantifies the bug detectability of a unit test $\unittest$ 
when applied on a 
$\buggycode$ and is computed as the fraction of problems for which a $\failtrace$ is produced. The AR for the $i^{th}$ problem is computed as $AR_i = \mathbf{1}$ if $\failtrace$, otherwise $\mathbf{0}$. The total attack rate is computed as $AR = \frac{\sum_{i = 1}^N AR_i}{N}$, where N is the total number of problems. A higher AR indicates the effectiveness of unit tests in 
bug detection. 


\smallskip

\noindent {\bf Divergent Rate (DR).} Divergent Rate quantifies how different/divergent a failure trace $\failtrace$ is as compared to a passing trace $\passtrace$ when the same unit test $\unittest$ is applied on a likely buggy design $\buggycode$ and an oracle $\oracle$, respectively. DR is measured as $DR = \frac{\sum_{i = 1}^{n}|\passtrace[i] \neq \failtrace[i]|}{n}$, where $||\unittest|| = n$. A higher DR value indicates a significant amount of 
mismatch 
between $\passtrace$ and $\failtrace$, 
yielding an 
informative failing trace for debugging. 
\bem{Note} while AR measures bug detectability, DR measures bug debuggability.



\smallskip

\noindent {\bf Divergent Attack (DA).} Divergent Attack quantifies the bug detectability and debuggability of a unit test $\unittest$ in a unified way. Intuitively, a test should be capable of detecting a broad spectrum of bugs (\ie, sensitive to bugs) and generate a unique failure signature for a bug to aid debugging (\ie, specific to the effects of the bug). Consequently, we compute DA as $DA = DR \cap AR$, \ie, a unit test $\unittest$ will have higher DA {\em if} it has both high AR and DR.

\smallskip
\noindent \bem{Experimental Categories}: We primarily compare {\tt 0-shot} of NLS (denoted as $\promptzero$) and {\tt 0-shot} and {\tt 5-shot} of NLSC, denoted as $\iclzero$ and $\iclfive$, respectively for the VerilogEval dataset. For CVDP, we use the 68-problem agentic commercial code-generation subset (spanning five categories) for the frontier model comparison and 336 naturally buggy generations for the iterative-debugging study; for LLM4DV we use all five designs. We have used Synopsys\textsuperscript{\textregistered} VCS V-2023.12-1 to simulate SV designs ({\tt -sverilog} flag) 
and collect 
the coverage information ({\tt -cm} flag). 

\smallskip
\section{Experimental Results}\label{sec:exp_results}

In this Section, we evaluate the performance 
of the generated tests for bug detection and debugging using the buggy design benchmark. 


\subsection{Bug Sensitivity Analysis of Unit Tests}\label{sec:attack_rate_ana}
 
In this experiment, we evaluate {\em how sensitive the generated unit tests are $\wrt$ bug detection}. 
Toward that, for a given hardware design task, we generated unit tests from the $i^{th}$ buggy code, $i \in [1, 10]$ and applied it on 
BC01, $\ldots$, BC10.
The results show different behaviors across the two LLMs. For Gemini-2.5 Pro, $\promptzero$ is highly effective, consistently achieving AR values falling in the fifth bin ($> 80\%$). This performance remains stable in $\iclzero$ and $\iclfive$, with the exception of $\iclzero$, where the AR for Source BC09 against Target BC07 drops significantly to $<20\%$.
DeepSeek R1 is ineffective with $\promptzero$; AR distributions remain largely in the lowest quartile ($<20\%$), except for specific targets (BC03 and BC08) which show higher susceptibility. However, the $\iclzero$ and $\iclfive$ configurations drastically improve DeepSeek R1's performance, raising the AR to levels that are competitive with Gemini Pro ($>80\%$) across all targets.
This experiment demonstrates that contextual grounding is a critical requirement for DeepSeek R1 to overcome generation volatility and achieve competitive bug sensitivity.

\subsection{Bug Specificity and Debuggability Analysis of Unit Tests}\label{sec:divergence_rate_ana}

In this experiment, we evaluate {\em the goodness of the unit tests in generating a bug-specific, divergent failure signature}, jointly via DR (specificity) and DA (detectability $\cap$ debuggability). Table~\ref{tab:bin_dist_comb} shows the DR bin distribution for Gemini-2.5 Pro and DeepSeek R1 (DA coincides with DR in this dataset since AR $=1$ for nearly all source-target pairs); Figure~\ref{fig:da_clean} details the Gemini-2.5 Pro DA grid. The color of each cell represents the median DR, mapped to a scale where \textcolor{red}{red} and \textcolor{green}{green} indicate worse (0\%) and optimal (100\%) performance, respectively, with the bin containing the median rendered in a darker shade to jointly convey variance and mean efficacy.
\begin{table}[t]
\centering
{%
\small
\setlength{\tabcolsep}{1pt}
\begin{tabular}{lcccccc}
\toprule
 & \multicolumn{3}{c}{Gemini Pro} & \multicolumn{3}{c}{DeepSeek R1} \\
\cmidrule(lr){2-4}\cmidrule(lr){5-7}
Bin & \promptzero  & \iclzero  & \iclfive & \promptzero & \iclzero & \iclfive \\
\midrule
0--20   & \cellcolor{blue!7}7.4\%  & \cellcolor{blue!9}9.0\%  & \cellcolor{blue!9}9.1\%  & \cellcolor{blue!30}30.3\% & \cellcolor{blue!4}3.6\%  & \cellcolor{blue!5}4.9\%  \\
20--40  & \cellcolor{blue!9}9.3\%  & \cellcolor{blue!10}9.8\%  & \cellcolor{blue!5}4.5\%  & \cellcolor{blue!6}6.0\%  & \cellcolor{blue!19}19.4\% & \cellcolor{blue!14}14.0\% \\
40--60  & \cellcolor{blue!35}35.2\% & \cellcolor{blue!7}6.6\%  & \cellcolor{blue!21}20.9\% & \cellcolor{blue!4}3.6\%  & \cellcolor{blue!31}30.9\% & \cellcolor{blue!29}28.7\% \\
60--80  & \cellcolor{blue!19}18.5\% & \cellcolor{blue!11}10.7\% & \cellcolor{blue!5}4.5\%  & \cellcolor{blue!14}13.8\% & \cellcolor{blue!25}24.5\% & \cellcolor{blue!34}34.1\% \\
80--100 & \cellcolor{blue!30}29.6\% & \cellcolor{blue!24}23.9\% & \cellcolor{blue!31}30.9\% & \cellcolor{blue!26}26.3\% & \cellcolor{blue!22}21.6\% & \cellcolor{blue!48}48.3\% \\
\bottomrule
\end{tabular}}
\caption{Number of source-target BC pairs (out of 100) whose median DR across problems falls in each bin, for Gemini Pro and DeepSeek R1 across \promptzero/\iclzero/\iclfive, for combinational 
designs. 
}
\label{tab:bin_dist_comb}
\end{table}
For $\promptzero$, Gemini-2.5 Pro performs the worst, with distributions spreading evenly across intermediate bins. It demonstrates consistency across $\iclzero$ and $\iclfive$, achieving up to 100\% (average 90\%) DR for BC08 -- BC10, though $\iclzero$ shows a specific vulnerability for BC02 and BC07: DA likewise stays high (80--100\%) for BC08--BC10 but drops to intermediate bins for BC01--BC07 despite non-zero AR, showing that successful attacks do not {\em guarantee strongly divergent failure traces} and that specific bug types may act as adversarial examples for unit test generation.
DeepSeek R1 shows a {\em high sensitivity} to the unit test generation strategy: $\promptzero$ is concentrated in Bin 1 ($0\%-20\%$), indicating a near-total inability to generate divergent traces without context; $\iclzero$ shows high variance, struggling to match Gemini-2.5; and $\iclfive$ consistently produces high DR/DA scores for complex bug targets, particularly BC08--BC10, demonstrating that few-shot guidance enables the model to detect bugs and propagate their effects through multiple simulation cycles.
\bem{This experiment shows that DeepSeek R1 heavily relies on few-shot examples to reason and achieve competitive performance, and that generating a unit test for real-world debugging requires creativity that can be challenging for state-of-the-art LLMs}.
\begin{figure}[t]
\centering
\begin{tikzpicture}[scale=0.5]
\definecolor{heatA}{HTML}{0F6E56}
\foreach \row [count=\ry] in {%
 {81,41,50,47,44,50,51,92,92,96},%
 {59,71,54,50,41,60,59,92,93,96},%
 {60,33,83,50,42,52,48,92,90,96},%
 {50,42,51,77,36,58,55,92,92,96},%
 {60,41,51,50,74,57,56,92,90,96},%
 {53,45,54,48,42,92,50,92,89,96},%
 {59,38,50,50,42,57,88,88,92,95},%
 {50,42,58,50,42,55,51,92,88,96},%
 {50,38,50,46,39,55,52,87,92,95},%
 {50,41,50,44,42,52,50,84,92,96}%
}{%
 \foreach \v [count=\cx] in \row {%
   \pgfmathtruncatemacro{\pc}{\v}%
   \ifnum\pc>55\relax\def\tcol{white}\else\def\tcol{heatA}\fi
   \fill[heatA!\pc!white] (\cx-1,10-\ry) rectangle (\cx,11-\ry);
   \node[text=\tcol,font=\tiny] at (\cx-0.5,10.5-\ry) {\v};
 }%
}
\draw[white,line width=0.6pt] (0,0) grid (10,10);
\foreach \i in {1,...,10}{%
  \pgfmathtruncatemacro{\n}{\i}%
  \node[rotate=90,anchor=east,font=\tiny] at (\i-0.5,-0.1) {BC\ifnum\n<10 0\fi\n};
  \node[anchor=east,font=\tiny] at (-0.05,10.5-\i) {BC\ifnum\n<10 0\fi\n};
}
\node[font=\scriptsize] at (5,-2.05) {Target buggy code};
\node[rotate=90,font=\scriptsize] at (-1.75,5) {Source buggy code};
\foreach \k in {0,...,19}{%
  \pgfmathtruncatemacro{\cp}{\k*5+2}%
  \fill[heatA!\cp!white] (10.7,\k*0.5) rectangle (11.2,\k*0.5+0.5);
}
\draw[black!45,line width=0.3pt] (10.7,0) rectangle (11.2,10);
\foreach \t/\lab in {0/0,5/50,10/100}{\node[anchor=west,font=\tiny] at (11.28,\t) {\lab};}
\node[rotate=-90,font=\tiny] at (12.25,5) {median DA};
\end{tikzpicture}
\caption{\textbf{Debuggability (DA) of $\pname$'{}s unit tests (Gemini-2.5 Pro).} 
DA over the $10\times10$ Source$\times$Target buggy-code grid for the NLSC@5 configuration for the combinational designs. Number denotes DA for different sets of buggy codes. 
}
\label{fig:da_clean}
\end{figure}

\subsection{Effectiveness of Iterative Debugging}
In this experiment, we evaluate {\em effectiveness of $\pname$'s iterative debugging process by leveraging the generated unit tests, failure trace $\failtrace$, and simulation feedback}.
Table~\ref{fig:debug_results} summarizes the debugging pass rate (mean$\pm$std) for the combinational and sequential designs, respectively, on $\promptzero$, $\iclzero$, and $\iclfive$.
For $\promptzero$, Gemini-2.5 Pro exhibits high success rates (median $\approx 0.85$ for combinational and $\approx 0.73$ for sequential), while DeepSeek R1 remains highly unstable, with a wide, dispersed distribution (median $\approx 0.5$) that indicates a lack of consistent reasoning when forced to debug with unit tests generated via $\promptzero$.
In $\iclzero$, both models improve as the availability of source code acts as a critical grounding mechanism for open-source models, and DeepSeek R1 achieves the largest relative gains (medians between 0.7 and 0.8). This 
suggests that while these models lack the intuition to generate high-quality unit tests in $\promptzero$, their reasoning engines are fully capable of analyzing specific bugs once the design circuit is exposed.
$\iclfive$ provides a universal, though more incremental, refinement, pushing median success rates for all models above 0.75 for combinational designs and closing the gap between proprietary and open-source models; Gemini-2.5 Pro retains a consistent edge, with tighter distributions near the upper bound compared to the larger variance in DeepSeek R1.
\begin{table}[t]
\centering
\begin{subtable}[b]{0.48\textwidth}
\centering
\small
\setlength{\tabcolsep}{3pt}
\begin{tabular}{lccc}
\toprule
Model & NLS@0 & NLSC@0 & NLSC@5 \\
\midrule
DeepSeek R1 & 0.49$\pm$0.05 & 0.76$\pm$0.03 & \cellcolor{green!20}0.78$\pm$0.06 \\
Qwen 32B & 0.13$\pm$0.02 & 0.72$\pm$0.04 & \cellcolor{green!20}0.75$\pm$0.07 \\
Gemini Pro & 0.85$\pm$0.02 & 0.90$\pm$0.02 & \cellcolor{green!20}0.93$\pm$0.02 \\
Gemini Flash & 0.84$\pm$0.03 & 0.87$\pm$0.02 & \cellcolor{green!20}0.89$\pm$0.03 \\
\bottomrule
\end{tabular}
\caption{}
\label{fig:debug_results_comb}
\end{subtable}
\hfill
\begin{subtable}[b]{0.48\textwidth}
\centering
\small
\setlength{\tabcolsep}{3pt}
\begin{tabular}{lccc}
\toprule
Model & NLS@0 & NLSC@0 & NLSC@5 \\
\midrule
DeepSeek R1 & 0.40$\pm$0.06 & 0.69 $\pm$0.06 & \cellcolor{green!20}0.72$\pm$0.07 \\
Qwen 32B & 0.25$\pm$0.04 & 0.57$\pm$0.05 & \cellcolor{green!20}0.69$\pm$0.06 \\
Gemini Pro & 0.73$\pm$0.03 & 0.84$\pm$0.03 & \cellcolor{green!20}0.85$\pm$0.01 \\
Gemini Flash & 0.81$\pm$0.04 & 0.81$\pm$0.03 & \cellcolor{green!20}0.82$\pm$0.03  \\
\bottomrule
\end{tabular}
\caption{}
\label{fig:debug_results_seq}
\end{subtable}
\caption{(\subref{fig:debug_results_comb}) \textbf{Debugging Pass Rate (mean$\pm$std) of $\promptzero$, $\iclzero$, and $\iclfive$ for various LLMs for buggy combinational circuit design codes, }
(\subref{fig:debug_results_seq}) \textbf{Debugging Pass Rate for buggy sequential circuit design codes}.
}
\label{fig:debug_results}
\end{table}

Overall, the performance of the models across each $\generator$ is lower for sequential designs than for combinational designs, reflecting the added complexity of state-dependent behavior in sequential circuits.
\bem{This experiment shows that while the current LLMs are effective in reasoning for debugging, they struggle to debug buggy designs without additional information}.

\subsection{Comparison with LLM4DV}\label{sec:llmdv}
In this experiment, we evaluate 
unit tests of $\pname$ with respect to the tests generated by LLM4DV in the context of detection, localization, and debuggability. We hypothesize that \pname{}'s simulation-grounded unit tests perform better than the LLM4DV baseline on 
AR as well as 
DR and that this is not an artifact of one favorable design but holds across structurally heterogeneous hardware.
To test this, we evaluate five designs of increasing structural diversity Stride Detector (SD), Ibex Decoder (ID), Ibex CPU via the decoder (IC), Async FIFO and SDRAM Controller injecting ten representative mutants per design and maintaining the number of stimulus sets fixed per design ($74$ for SD, $76$ for ID, $15$ for IC, $45$ each for FIFO and SDRAM). We apply both LLM4DV and \pname{} generated stimuli and report AR, DR, DA, and Mutation Score~(MS), as shown in Table~\ref{tab:llm4dv_laude}. We observe that \pname{} outperforms LLM4DV on every design and every metric, without regressions (\eg, SD: $\text{AR}=0.804 \to 0.891$; Async FIFO reaches a perfect $\text{AR}=\text{DA}=1.000$ with $\text{MS}=10/10$; SDRAM DR: $0.519 \to 0.573$). 
We next hypothesize that bug detectability is governed primarily by the nature of the injected fault rather than by the model or prompting strategy, i.e., that detectability varies sharply mutant-to-mutant even under a fixed generator. To test this, we decompose the SD and ID results into their ten individual mutants~(M1-M10) and examine the per-mutant AR, DR, and DA. 
We observe a wide spread: some mutants are completely undetectable - M7~(\texttt{stride\_2} wrong slot) in SD, and M3~(SLT/SLTU swapped) and M9~(Load \texttt{data\_type} byte/word swap) in ID all yield $\text{AR}=\text{DR}=\text{DA}=0.000$ while others are highly informative, with M2~(confidence decrement off-by-one) in SD reaching $\text{DR}=0.950$ and M8~(\texttt{rf\_we\_o} always~$0$) in ID reaching $\text{DR}=0.788$. From this we conclude that 
detectability is dominated by fault semantics 
and that per-mutant reporting is necessary to expose the 
coverage gap. \pname{} helps to obtain unit tests focused on fault semantics that increase coverage.

\begin{table}
\centering
{%
\small
\begin{tabular}{L{1cm}L{1cm}C{1cm}C{0.6cm}C{0.6cm}C{0.6cm}C{0.6cm}}
\toprule
\textbf{Design} & \textbf{Stimuli} & \textbf{Stimuli} &
\textbf{AR} & \textbf{DR} & \textbf{DA} & \textbf{MS} \\
& \textbf{Type} & \textbf{Sets} &
&  & & \\

\midrule
\multirow{2}{*}{{\bf SD}}
  & LLM4DV      & 74 & 0.804 & 0.226 & 0.804 & 0.9 \\
  & $\pname$              & 74 & \cellcolor{green!20}0.891 & \cellcolor{green!20}0.261 & \cellcolor{green!20}0.891 & \cellcolor{green!20}1 \\
\midrule
\multirow{2}{*}{{\bf ID}}
  & LLM4DV   & 76 & 0.566 & 0.178 & 0.566 & 1 \\
  & $\pname$              & 76 & \cellcolor{green!20}0.743 & \cellcolor{green!20}0.215 & \cellcolor{green!20}0.743 & \cellcolor{green!20} 1 \\
\midrule
\multirow{2}{*}{{\bf IC}}
  & LLM4DV   & 15 & 0.713 & 0.199 & 0.713 & 0.8 \\
  & $\pname$              & 15 & \cellcolor{green!20}0.867 & \cellcolor{green!20}0.241 & \cellcolor{green!20}0.867 & \cellcolor{green!20}0.9 \\
\midrule
\multirow{2}{*}{{\bf AFIFO}}
  & LLM4DV   &  45 & 0.900 & 0.565 & 0.900 & 0.9 \\
  & $\pname$              &  45 & \cellcolor{green!20}1.000 & \cellcolor{green!20}0.612 & \cellcolor{green!20}1.000 & \cellcolor{green!20}1 \\
\midrule
\multirow{2}{*}{{\bf SC}}
  & LLM4DV  &  45 & 0.700 & 0.519 & 0.700 & 0.7 \\
  & $\pname$              &  45 & \cellcolor{green!20}0.850 & \cellcolor{green!20}0.573 & \cellcolor{green!20}0.850 & \cellcolor{green!20}0.9 \\
\bottomrule
\end{tabular}
}
\caption{LLM4DV $\pname$ mutation-testing results.
  AR, DR, DA and MS for {\bf Designs}: {\bf SD}: Stride Detector. {\bf ID}: Ibex Decoder. {\bf IC}: Ibex CPU via ID. {\bf AFIFO}: Async FIFO. {\bf SC}: SDRAM Controller.}
\label{tab:llm4dv_laude}
\end{table}

\subsection{Comparison with CVDP}\label{sec:cvdp}
We evaluated two leading models, Gemini~2.5 and Claude Sonnet~4.6, on the $68$ agentic code-generation problems of the commercial CVDP dataset, reporting Problem Pass Rate~(PPR, where a problem counts only if all its tests pass) and Test Pass Rate~(TPR, the fraction of individual testbench stimuli passed), broken down by difficulty (Easy, Medium, Hard) in Table~\ref{tab:cvdp_overall}. Claude Sonnet~4.6 slightly outperforms Gemini~2.5 overall ($47.76\%$ vs.\ $45.59\%$ PPR; $52.63\%$ vs.\ $50.65\%$ TPR), identical on Easy and Hard, with Claude's only edge on Medium ($66.67\%$ vs.\ $62.16\%$). Aggregate scores are thus nearly indistinguishable and Hard pass rates are low for both models, confirming that overall PPR/TPR alone is too coarse to separate frontier models, motivating the finer breakdowns below.
We hypothesize that the difficulty a model faces is driven less by the nominal difficulty label and more by the design family of the problem, i.e., that structural RTL and sequential/cryptographic designs sit at opposite ends of tractability regardless of model.
\begin{table}[htbp]

{%
\small
\begin{tabular}{C{1.2cm}C{1.2cm}C{1.2cm}C{1.5cm}C{1.2cm}}
\toprule
\textbf{Category} & \textbf{Problems} &
\textbf{Gemini~2.5} & \textbf{Claude Sonnet~4.6} &
\textbf{Difficulty} \\
\midrule
cid003 & 6  & 50.00\% & 50.00\% & H  \\
cid005 & 7  & 57.14\% & 57.14\% & H  \\
cid012 & 16 & 100.00\% & 100.00\% & E/M/H \\
cid013 & 20 & 10.00\% & 15.79\% & E/M/H \\
cid014 & 28 & 50.00\% & 50.00\% & E/M/H \\
\midrule
\textbf{Overall} & \textbf{68/68} & \textbf{45.59\%} & \textbf{47.76\%} & -- \\
\bottomrule
\end{tabular}
}
\caption{CVDP agentic code-generation 
  by category. Category IDs map to design families:
  cid003 = multi-component hard designs (8b10b codec, DRAM controller);
  cid005 = algorithmic hard designs (UART, Hamming distance);
  cid012 = structural RTL (FIFO, cache, RAM, BCD adder);
  cid013 = sequential (64b66b codec, ALU, BCD-excess-3);
  cid014 = complex IP (AES, AXI4-Lite, barrel shifter). {\bf E}: Easy. {\bf M}: Medium. {\bf H}: Hard.}
\centering
\label{tab:cvdp_category}
\end{table}

We apply $\pname$'s debugging loop (up to five rounds) to the buggy generations to track the cumulative pass rate after each round, shown in Figure~\ref{fig:cvdp_flow}. We observe that both models improve monotonically: Claude Sonnet~4.6 rises from $31\%$ (initial generation) to $58\%$ after five rounds ($+27$pp), while Gemini~2.5 rises from $28\%$ to $49\%$ ($+21$pp), with both curves showing diminishing per-round gains 
as the easier bugs are resolved first. 
\begin{figure}[t]
\centering
\begin{tikzpicture}[x=1.14cm, y=0.085cm]
\definecolor{claudeC}{HTML}{E8743B}
\definecolor{geminiC}{HTML}{1F77B4}

\draw[black!60, thin] (-0.3,20) rectangle (5.3,65);

\foreach \y in {20,25,...,65}{
  \draw[black!15] (-0.3,\y) -- (5.3,\y);
  \node[anchor=east, font=\tiny] at (-0.4,\y) {\y};
}
\foreach \x/\lab in {0/Initial,1/Round 1,2/Round 2,3/Round 3,4/Round 4,5/Round 5}{
  \node[anchor=north, font=\tiny] at (\x,19) {\lab};
}
\node[font=\small] at (2.5,69) {\textbf{Cumulative pass rate over debugging rounds}};
\node[rotate=90, font=\tiny] at (-1.7,42.5) {Cumulative pass rate (\%)};
\node[font=\tiny] at (2.5,14.5) {Debugging round};

\draw[claudeC, thick] (0,31) -- (1,41) -- (2,48) -- (3,53) -- (4,56) -- (5,58);
\foreach \x/\y in {0/31,1/41,2/48,3/53,4/56,5/58}{
  \node[circle, fill=claudeC, minimum size=4pt, inner sep=0pt] at (\x,\y) {};
  \node[claudeC, anchor=south, font=\tiny] at (\x,\y+2.2) {\y\%};
}
\draw[geminiC, thick,] (0,28) -- (1,36) -- (2,42) -- (3,46) -- (4,48) -- (5,49);
\foreach \x/\y in {0/28,1/36,2/42,3/46,4/48,5/49}{
  \node[rectangle, fill=geminiC, minimum width=3.2pt, minimum height=3.2pt, inner sep=0pt] at (\x,\y) {};
  \node[geminiC, anchor=north, font=\tiny] at (\x,\y-2.2) {\y\%};
}
\draw[claudeC, thick] (0.1,61.5) -- (0.55,61.5);
\node[circle, fill=claudeC, minimum size=4pt, inner sep=0pt] at (0.325,61.5) {};
\node[anchor=west, font=\tiny] at (0.65,61.5) {Claude};
\draw[geminiC, thick,] (0.1,58.5) -- (0.55,58.5);
\node[rectangle, fill=geminiC, minimum width=3.2pt, minimum height=3.2pt, inner sep=0pt] at (0.325,58.5) {};
\node[anchor=west, font=\tiny] at (0.65,58.5) {Gemini};
\end{tikzpicture}
\caption{
     {{\bf CVDP pass rate increasing as the number of debugging rounds increases. Starting from each model's generation, both Claude and Gemini improve monotonically as \pname's debugging loop is applied.
     }}}
\label{fig:cvdp_flow}
\vspace{-5mm}
\end{figure}
 To test this, we break down the PPR in the five design categories as 
shown in Table~\ref{tab:cvdp_category}.
We observe a stark split: both models reach a perfect $100\%$ on cid012 (structural RTL such as FIFO, cache, and RAM), yet collapse on cid013 (sequential and cryptographic designs) to $10\%$ (Gemini) and $15.79\%$ (Claude), while cid003, cid005, and cid014 are tied at $50\%$--$57\%$.
\begin{table}[htbp]
{%
\small
\begin{tabular}{L{1.25cm}|C{0.25cm}|C{0.8cm}|C{0.8cm}|C{0.8cm}|C{0.8cm}|C{0.8cm}}
\hline
\textbf{Model} &
\textbf{P} &
\textbf{PPR} &
\textbf{TPR} &
\multicolumn{3}{c}{\textbf{PPR by difficulty}} \\
\cline{5-7}
& & & & E & M & H \\
\hline
Gemini~2.5        & 68 & 45.59\% & 50.65\% & 40.00\% & 62.16\% & 12.50\% \\
Sonnet~4.6 & 68 & 47.76\% & 52.63\% & 40.00\% & 66.67\% & 12.50\% \\
\hline
\end{tabular}
}
\caption{CVDP agentic code-generation 
  results by model.
  \textbf{TPR} = Test Pass Rate; \textbf{PPR} = Problem Pass Rate. {\bf P}: \# of Problem. {\bf E}: Easy. {\bf M}: Medium. {\bf H}: Hard.}
\centering
\label{tab:cvdp_overall}
\end{table}


We observe that Claude's advantage is concentrated on Hard problems in cid003 ($3/6=50\%$ vs.\ $0/3=0\%$) and cid005 ($4/7=57\%$ vs.\ $0/3=0\%$), where Gemini fails outright, while both models score $0\%$ on Hard cid014 and on Easy cid013. The takeaway is that frontier models are not uniformly ranked: Claude is decisively stronger on 
complex hard designs, yet sequential 
designs remain universally difficult. 
To summarize separation, we compute percentage-point deltas $\Delta = \text{Claude Sonnet~4.6} - \text{Gemini~2.5}$ across difficulty, category, and category~$\times$~difficulty. 
Claude leads by $+2.17$~pp PPR and $+1.98$~pp TPR overall, entirely from Medium difficulty ($+4.51$~pp; Easy and Hard at $\pm 0.00$) and from cid013 ($+5.79$~pp); at the finest granularity the largest gains are Hard cid003 ($+50.00$~pp) and cid005 ($+57.14$~pp), while the apparent $-66.67$~pp on Hard cid013 reflects Gemini's $1/1$ versus Claude's $1/3$. 
This experiment shows that frontier code-generation models remain unevenly matched on complex, naturally occurring hardware bugs, with the gap concentrated in sequential 
designs.

\section{Related Work}\label{sec:related_work}


Traditional hardware 
verification techniques 
are 
time-consuming, error-prone, and difficult to scale. Recently, the 
success of gen-AI, \eg, LLMs in 
scientific 
domains has propelled researchers to leverage LLMs reasoning 
for hardware design and verification tasks.


\smallskip 

\noindent \bem{LLMs for Code Generation.} The majority of recent efforts have focused on applying LLMs to {\em HDL Code Generation} 
~\cite {zhao2025arXiv,sun2025arxiv, vijayaraghavan2024mlcad, zhang2024lad, chang2025arxiv, liu2023iccad,pinckney2025arxiv, al_amin2025meco}. 
However, these works 
do not encompass the 
testing and debugging loops required for robust hardware verification. 

\noindent \bem{LLMs for Hardware Verification.} 
Recent work~\cite{blocklove2024lad, zhang2025arxiv, qayyum2024ats} benchmarked LLMs 
for hardware design, testbench generation, and hardware verification. However, all these methods require extensive manual prompting 
and careful integration due to data scarcity and domain complexity. $\pname$ 
leverages LLMs' 
reasoning abilities and 
advanced prompting strategies 
to fully automate unit test generation 
and facilitate bug localization. The closest baseline is LLM4DV~\cite{zhang2025arxiv}, which generates test stimuli without simulation grounding; $\pname$ couples LLM reasoning with simulation feedback and, as we show, outperforms LLM4DV on every design and metric. 

\section{Limitations}\label{sec:limitations}

We identify the following limitations of this work in terms of the dataset and the evaluation methodology.

\bem{Quantitative Evaluation}: In this work, we focus primarily on unit test generation and debugging without quantifying and ranking the subtleties of the captured design behavior. 
There is scope to include such rankings in the few-shot examples in $\baseline$, $\prompt$, and $\icl$ and evaluate the LLM's capability to 
rank quality of the debugged 
code. 

\bem{Modeling}: In this paper, we assessed the debugging capabilities of various language models. In future work, it will be interesting to fine-tune language models for unit test generation, code generation, debugging, and evaluation of their performance using $\pname$. Additionally, it would be interesting to develop language models that can predict the design behavior across clock cycles to eliminate the need for the oracle $\oracle$ in the $\pname$ flow.


\section{Conclusion}\label{sec:conclusion}

$\pname$ 
provides a scalable and automated solution for localizing, diagnosing, and debugging 
functional errors in the hardware design source code using unit tests. The experimental results on a commonly used benchmark and varied functional bugs demonstrate its efficacy in detecting and debugging functional errors. 
$\pname$ 
demonstrated that while the current closed- 
and open-source LLMs can reason on complex hardware verification tasks, there is 
broad scope for improvement.

\section{Appendix}

This appendix provides the implementation details and additional experiments that complement the Experimental Results section of the main paper. The full debugging prompt template and the ten bug types (BC01-BC10) used to construct the buggy-design benchmark are added in this appendix. 
The Interpretation of Violin Plots section complements both sets of results with a violin-plot analysis of the debugging success-rate distributions for all four LLMs. Finally, the LLM4DV Per-Mutant Breakdown section reports the per-mutant breakdown underlying the LLM4DV comparison, and the CVDP section reports the percentage-point breakdowns underlying the CVDP comparison.

\subsection{Debugging Prompt Template}\label{sec:debug_prompt_appendix}
Figure~\ref{fig:debug_prompt} provides the full debugging prompt template referenced in the ``Unit Test-Assisted Debugging Technique'' subsection of the main paper. The ten types of bugs (BC01-BC10) used to construct the buggy-design benchmark are listed in Table~\ref{tab:bug_types}. The reference Verilog code used to construct these buggy variants is given in Figure~\ref{fig:refmodule}.

\subsection{Sequential Bug Specificity}\label{sec:bin_dist_seq_appendix}

\begin{figure}
    \begin{tabular}{c}
\begin{lstlisting}[style=customtext]
You are an expert Verilog debugging engineer with deep expertise in logic circuit design and debugging. Your mission is to meticulously analyze comprehensive simulation feedback and systematically fix bugs in faulty Verilog implementations.

DEBUGGING SESSION - ITERATION <NUMBER>
==========================================
- Task: <TASK_DESCRIPTION>
- Module Interface: <MODULE_INTERFACE>
- Input ports: <INPUT_LIST>
- Output ports: <OUTPUT_LIST>
- Current Buggy Code Under Analysis: <SV_CODE>
- Simulation Artifacts: [Test-Vector Count | Quality Metric]
- Simulation Results: [Test Status | Signal Values | Signal Value Mismatch Details]
- Debugging Strategies: [Strategy 1 | Strategy 2| ... | Strategy N]
\end{lstlisting}
    \end{tabular}
    \caption{\textbf{Example prompt to guide an LLM for debugging}.  
The prompt 
specifies the task description ($\mathcal{L}$), likely buggy design source code ($\buggycode$), and formatting requirements, 
encouraging the generation of corrected code while preserving 
port ordering and design notation. {\bf Debugging Strategies} include {\em Clock Domain Analysis}, {\em Reset Logic Verification}, {\em State Machine Analysis}, {\em Edge Detection}, {\em Data Path Synchronization} among others.
}
    \label{fig:debug_prompt}
    \vspace{-5mm}
\end{figure} 

\begin{figure}[t]
    \centering
\begin{lstlisting}[language=Verilog,
             framexleftmargin=0.2pt,
             framexrightmargin=0.2pt,
             basicstyle=\ttfamily\footnotesize,
             ]
module RefModule (
  input clk,
  input [7:0] in,
  input reset,
  output done
);

  parameter BYTE1=0, BYTE2=1, BYTE3=2, DONE=3;
  reg [1:0] state;
  reg [1:0] next;

  wire in3 = in[3];

  always_comb begin
    case (state)
      BYTE1: next = in3 ? BYTE2 : BYTE1;
      BYTE2: next = BYTE3;
      BYTE3: next = DONE;
      DONE: next = in3 ? BYTE2 : BYTE1;
    endcase
  end

  always @(posedge clk) begin
    if (reset) state <= BYTE1;
      else state <= next;
  end

  assign done = (state==DONE);

endmodule
\end{lstlisting}
    \caption{Verilog code of a byte-reassembly FSM (\texttt{RefModule}) for which we have included the list of bugs.}
    \label{fig:refmodule}
\end{figure}

\begin{table*}
\centering
\small
\setlength{\tabcolsep}{1pt}
\begin{tabular}{cll}
\toprule
\textbf{ID} & \textbf{Error Category} & \textbf{Technical Description} \\
\midrule
BC01 & Incorrect Byte Ordering       & Reverses the order of 8-bit chunks. \\
BC02 & Off-by-one Bit Selection      & Overlapping selection where chunks share bits (e.g., [7:0] and [16:9]). \\
BC03 & Incorrect Chunk Size          & Changes granularity by swapping 4-bit nibbles instead of 8-bit bytes. \\
BC04 & Partial Chunk Inversion       & Inverts the internal bit order of a single chunk while others remain correct. \\
BC05 & Repeated Chunk                & Duplicates one 8-bit chunk and omits another, causing data loss. \\
BC06 & Zero Padding    & Omits a required chunk and replaces it with a constant \texttt{8'h00}. \\
BC07 & Unrelated Bit Permutation     & Implements a completely different pattern (e.g., 16-bit word swap). \\
BC08 & Constant Assignment          & Ignores input entirely and assigns a literal value (e.g., \texttt{32'hdeadbeef}). \\
BC09 & Off-by-one Bit Shift          & Performs a 1-bit left shift across the word, losing the MSB. \\
BC10 & Structural Bypass             & Defines internal wires but assigns the output directly to the input. \\
\bottomrule
\end{tabular}
\caption{One set of example for the ten injected bug types (BC01-BC10) used to construct the VerilogEval buggy-design suite.}
\label{tab:bug_types}
\end{table*}

This Section provides the sequential design companion to the combinational DR bin-distribution table in the ``Bug Specificity and Debuggability Analysis of Unit Tests'' subsection 
of the main paper. Table~\ref{tab:bin_dist_seq} reports, out of the 100 source-target buggy-code pairs, the number whose median DR falls in each of the five bins, for Gemini-2.5 Pro and DeepSeek R1 across $\promptzero$, $\iclzero$, and $\iclfive$ on the 74 sequential problems.

\begin{table}[t]
\centering
{%
\small
\setlength{\tabcolsep}{1pt}
\begin{tabular}{lcccccc}
\toprule
 & \multicolumn{3}{c}{Gemini Pro} & \multicolumn{3}{c}{DeepSeek R1} \\
\cmidrule(lr){2-4}\cmidrule(lr){5-7}
Bin & \promptzero & \iclzero & \iclfive & \promptzero & \iclzero & \iclfive\\
\midrule
0--20   & \cellcolor{blue!7}6.5\%  & \cellcolor{blue!30}29.5\% & \cellcolor{blue!23}23.3\% & \cellcolor{blue!6}5.6\%  & \cellcolor{blue!15}14.8\% & \cellcolor{blue!14}14.3\% \\
20--40  & \cellcolor{blue!29}28.8\% & \cellcolor{blue!13}13.2\% & \cellcolor{blue!10}9.7\%  & \cellcolor{blue!25}25.3\% & \cellcolor{blue!6}5.6\%  & \cellcolor{blue!18}18.2\% \\
40--60  & \cellcolor{blue!15}15.0\% & \cellcolor{blue!19}18.8\% & \cellcolor{blue!29}29.1\% & \cellcolor{blue!38}37.6\% & \cellcolor{blue!34}34.2\% & \cellcolor{blue!16}15.7\% \\
60--80  & \cellcolor{blue!35}35.3\% & \cellcolor{blue!29}28.6\% & \cellcolor{blue!22}21.8\% & \cellcolor{blue!19}19.1\% & \cellcolor{blue!29}28.6\% & \cellcolor{blue!20}20.0\% \\
80--100 & \cellcolor{blue!14}14.4\% & \cellcolor{blue!10}9.8\%  & \cellcolor{blue!16}16.0\% & \cellcolor{blue!12}12.4\% & \cellcolor{blue!17}16.8\% & \cellcolor{blue!32}31.8\% \\
\bottomrule
\end{tabular}}
\caption{Number of source-target BC pairs (out of 100) whose median DR across problems falls in each bin, for Gemini Pro and DeepSeek R1 across \promptzero/\iclzero/\iclfive, for sequential designs.}
\label{tab:bin_dist_seq}
\end{table}

\subsection{Debugging Success Rate Distribution Analysis}\label{sec:appendix_b}

This Section details the debugging success-rate distributions for all four LLMs using violin plots, which capture the full shape of model performance beyond simple summary statistics such as means or medians.

\subsubsection{Interpretation of Violin Plots}\label{sec:violin_interpretation}

The violin plots presented in Figure~\ref{fig:debug_results_violins_appendix} and Figure~\ref{fig:debug_results_violins_appendix_r} combine the features of a box plot with a kernel density estimation
. This visualization technique allows for a comprehensive assessment of the models' reliability and stability across different prompting strategies. The width of the violin at any given y-axis value represents the probability density of the success rate at that level. Wider sections indicate a higher frequency of data points (\ie, the model achieves this success rate more often), while narrower sections indicate rarity. The shape of the plot shows how the data is distributed. A bottom-heavy shape means most values are lower, while a top-heavy or wider upper part suggests consistently high performance.

\subsubsection{Effectiveness of Iterative Debugging using Violin Plots}\label{sec:violin_iterative_debugging}

We now analyze the effectiveness of $\pname$'s debugging process in fine-grained detail.

\begin{figure}
   \centering
    \begin{subfigure}[b]{0.1\columnwidth}
        \centering
        \includegraphics[scale=0.1,center
        ]{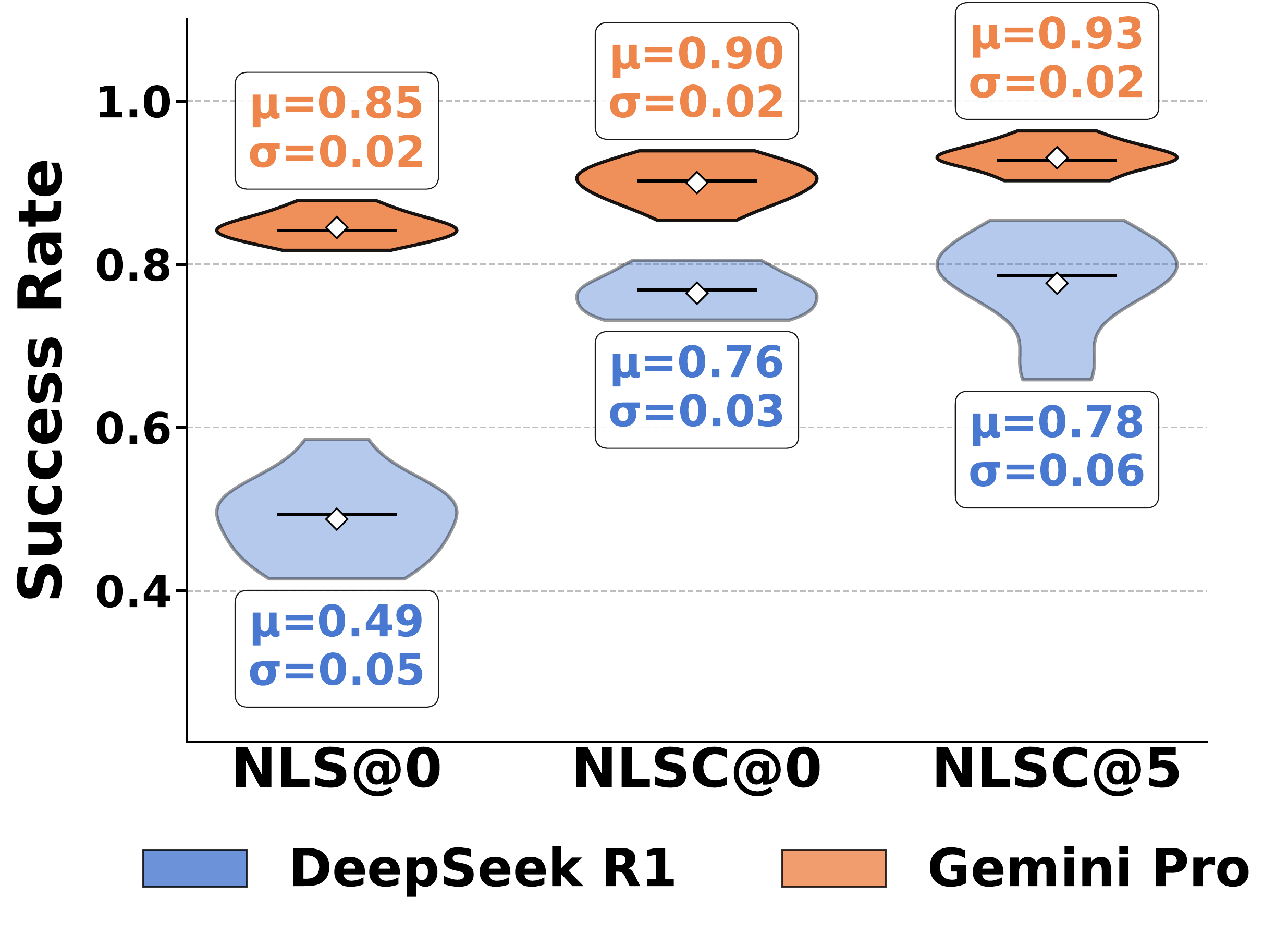}
    \caption{}
    \label{fig:debug_results_comb_violins_appendix}
    \end{subfigure}
    \hspace{32mm}
    \begin{subfigure}[b]{0.1\columnwidth}
        \centering
        \includegraphics[scale=0.1, center
        ]{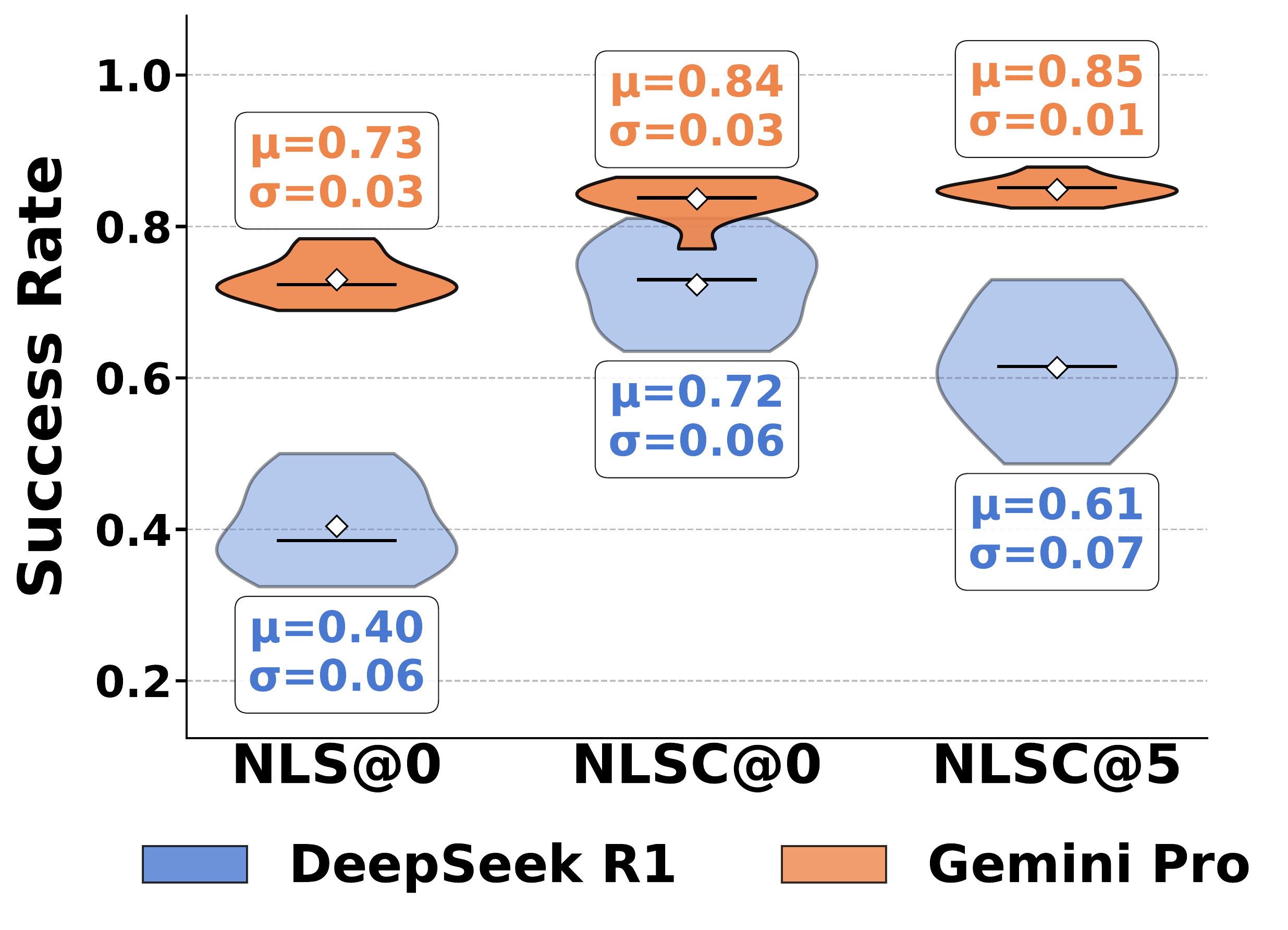}
    \caption{}
    \label{fig:debug_results_seq_violins_appendix}
    \end{subfigure}   
    \vspace{-4mm}
     \caption{(\subref{fig:debug_results_comb_violins_appendix}): \textbf{Debugging success rates of $\promptzero$, $\iclzero$, and $\iclfive$ for Gemini-2.5 Pro and DeepSeek R1 for buggy {\em combinational circuit} design codes}. 
    ~(\subref{fig:debug_results_seq_violins_appendix}): \textbf{Debugging success rates of $\promptzero$, $\iclzero$, and $\iclfive$ for Gemini 2.5 Pro and DeepSeek R1 
    for buggy {\em sequential circuit} design codes}. 
    }
    \label{fig:debug_results_violins_appendix}
\end{figure}

\paragraph{Gemini-2.5 Pro and DeepSeek R1.} Figure~\ref{fig:debug_results_violins_appendix} shows the violin plot distributions for the success rates for Gemini-2.5 Pro and DeepSeek R1 in the three configurations.

For combinational designs, we see that \textbf{Gemini-2.5 Pro} has top-heavy violin shapes across all configurations. These upper sections indicate a ceiling effect; the model consistently averages a high score and hits the maximum possible success rate with very few outliers. In contrast, \textbf{DeepSeek R1} in the $\promptzero$ configuration shows a flat, stretched shape, meaning its performance is effectively random.
Adding prompts ($\iclzero$ and $\iclfive$) compresses this shape into a tighter blob in the upper range, showing that structure improves the score and stabilizes the model's behavior.

For sequential designs, we see that \textbf{Gemini-2.5 Pro} remains consistent; the violins are compact and narrow, showing little variation between different runs. The interesting detail appears in \textbf{DeepSeek R1}. In the $\promptzero$ setting, the mass is concentrated at the bottom (consistent failure). While the $\iclzero$ prompt moves this mass clearly upward, the $\iclfive$ (few-shot) configuration results in a wider, fatter shape. This visible expansion in width indicates that while few-shot examples help some test cases, they introduce higher variance in others, making the model less predictable than with simple instructions.

\bem{The visual spread observed in this experiment shows that DeepSeek R1's results distribution suggests that the model gets ``noisy'' when given few-shot examples for complex state logic. In contrast, Gemini-2.5 Pro scales cleanly with more data at its disposal}.

\paragraph{Gemini-2.5 Flash and Qwen-2.5 Coder.} We extend our analysis to \textbf{Gemini-2.5 Flash} and \textbf{Qwen-2.5 Coder} using violin plots to visualize their stability and responsiveness to different prompting strategies.

\begin{figure}
   \centering
    \begin{subfigure}[b]{0.1\columnwidth}
        \centering
        \includegraphics[scale=0.1,center
        ]{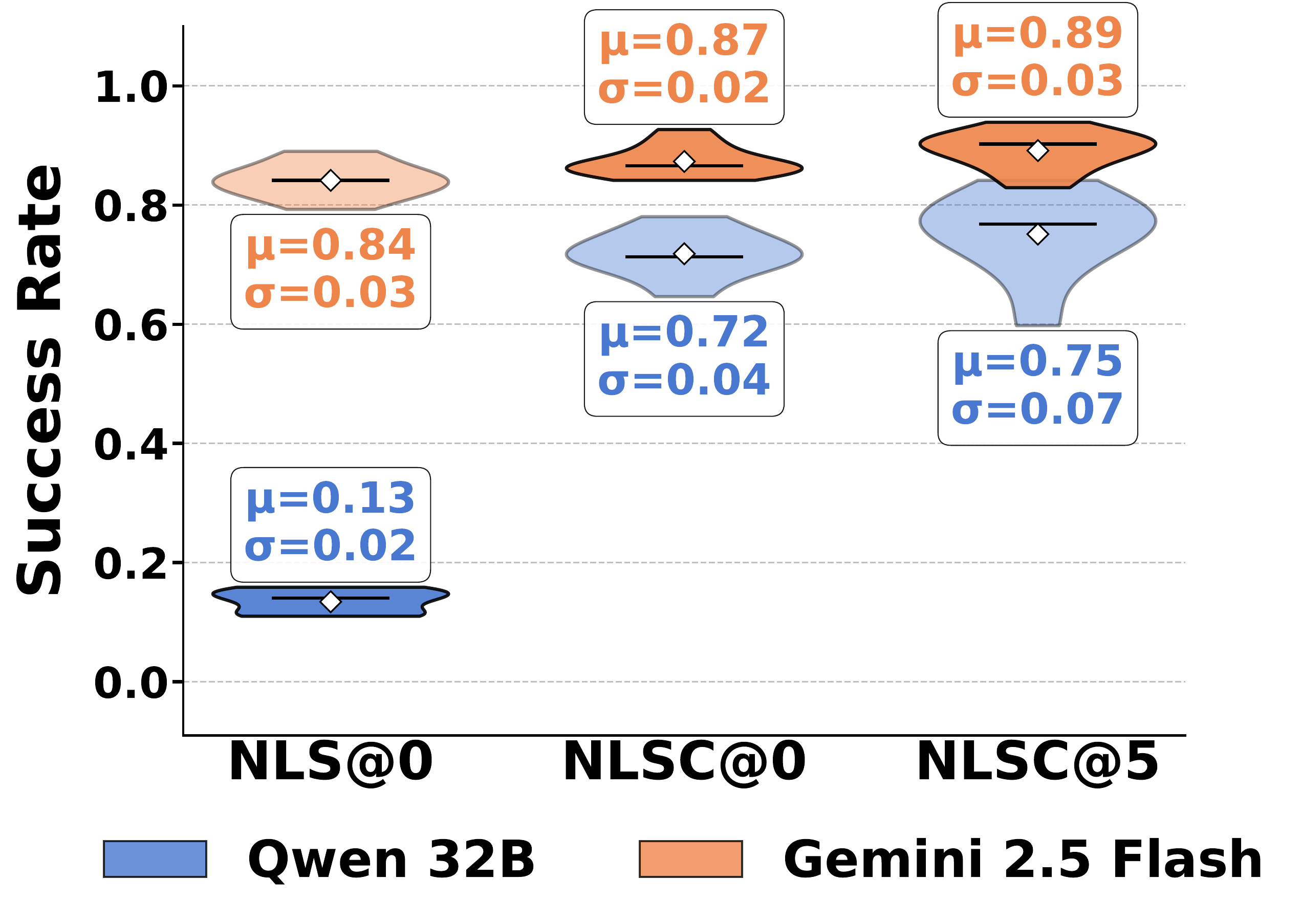}
    \caption{}
    \label{fig:debug_results_comb_violins_appendix_r}
    \end{subfigure}
    \hspace{32mm}
    \begin{subfigure}[b]{0.1\columnwidth}
        \centering
        \includegraphics[scale=0.1, center
        ]{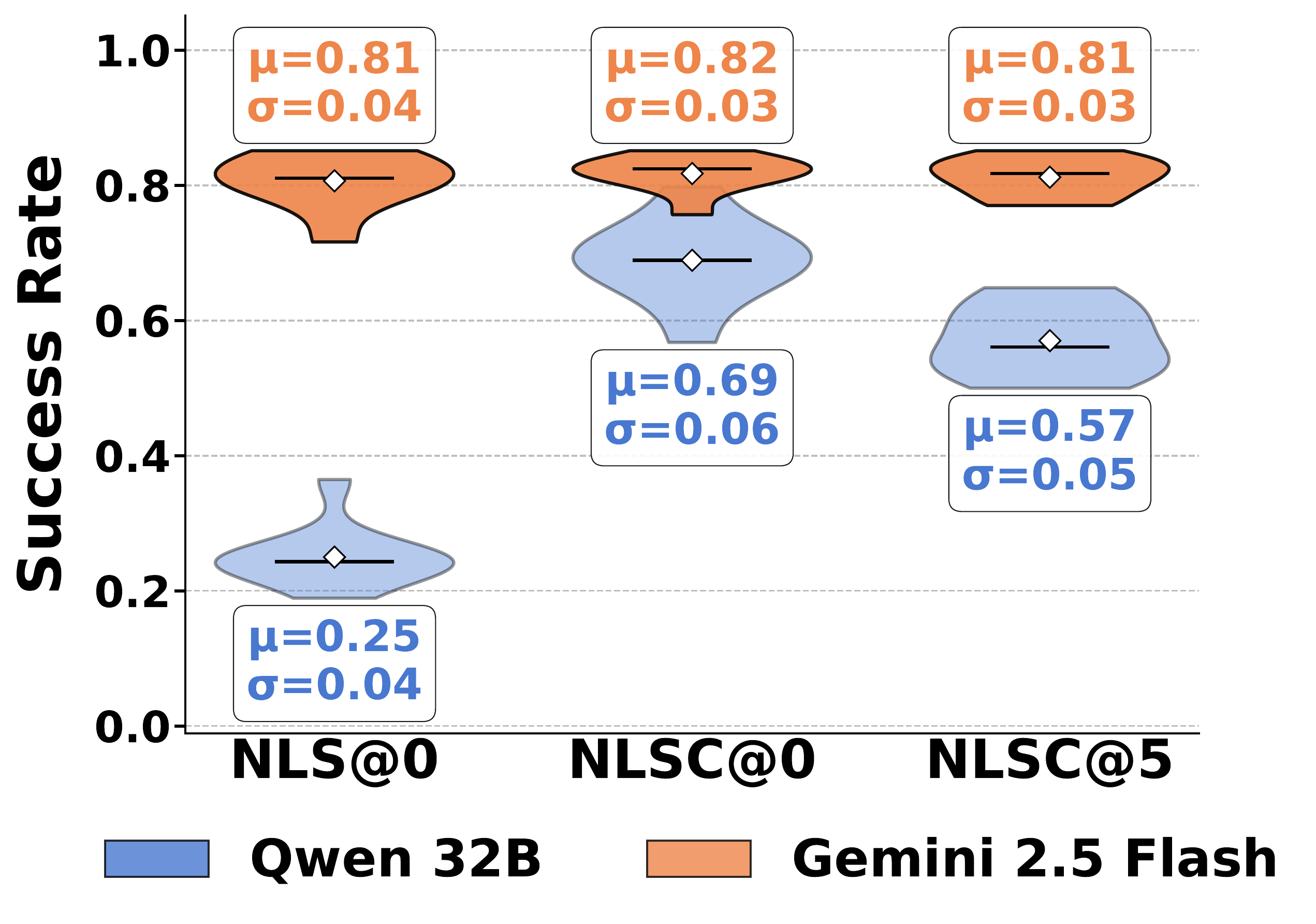}
    \caption{}
    \label{fig:debug_results_seq_violins_appendix_r}
    \end{subfigure}   
    \vspace{-4mm}
     \caption{(\subref{fig:debug_results_comb_violins_appendix_r}): \textbf{Debugging success rates of $\promptzero$, $\iclzero$, and $\iclfive$ for Gemini-2.5 Flash and Qwen-2.5 32B Coder 
     for buggy {\em combinational circuit} design codes}. 
    ~(\subref{fig:debug_results_seq_violins_appendix_r}): \textbf{Debugging success rates of $\promptzero$, $\iclzero$, and $\iclfive$ for Gemini-2.5 Flash and Qwen-2.5 32B Coder
    for buggy {\em sequential circuit} design codes}. 
    }
    \label{fig:debug_results_violins_appendix_r}
    \vspace{-5mm}
\end{figure}

Considering the combinational circuits, Figure~\ref{fig:debug_results_comb_violins_appendix_r} reveals a contrast in baseline capability. \textbf{Gemini-2.5 Flash} is remarkably consistent; even in the simple $\promptzero$ configuration, the distribution is concentrated high up ($\mu=0.84$), and adding context in $\iclfive$ only slightly tightens this top-heavy shape ($\mu=0.89$). It works well. In contrast, \textbf{Qwen-2.5 Coder} is effectively broken in the $\promptzero$ setting, with a flat distribution stuck at the bottom ($\mu=0.13$). However, the shape of the violin plot changes dramatically with the $\iclzero$ prompt, jumping massively to a median of $\mu=0.72$. This gap proves that Qwen possesses the logic to solve these circuits but is completely dependent on specific instructions to unlock them, whereas Gemini-2.5 Flash is self-sufficient.

For sequential designs, Figure~\ref{fig:debug_results_seq_violins_appendix_r} highlights a critical weakness in Qwen's handling of context. \textbf{Gemini-2.5 Flash} remains prompt-invariant; its violins are almost identical across all three settings ($\mu \approx 0.81$), showing that neither instructions nor few-shot examples significantly change its stable behavior. \textbf{Qwen-2.5 Coder}, however, shows a unique regression. While it improves from $\promptzero$ ($\mu=0.25$) to $\iclzero$ ($\mu=0.69$), adding the few-shot examples in $\iclfive$ actually hurts performance, dropping the median to $\mu=0.57$. The violin shape for NLSC becomes wider and shifts lower than NLS, indicating that for this model, the extra text from the few-shot examples acts as a distraction rather than a guide.

\bem{The violin plots expose a ``Contextual Distraction'' effect in Qwen-2.5 Coder: Gemini-2.5 Flash ignores prompt format and stays flat around $\mu=0.81$ regardless of configuration, whereas Qwen is highly sensitive improving from $\promptzero$ to $\iclzero$ but regressing under $\iclfive$ on sequential tasks, a 12\% drop suggesting that smaller open-source models can struggle to attend to relevant details once the prompt grows too long}.
\begin{table}[htbp]
\caption{LLM4DV $\pname$ per-mutant results for Stride Detector (SD) and Ibex Decoder (ID) using LLM-generated stimuli (averaged across all experiment-log stimulus sets).}
\centering
\label{tab:llm4dv_per_mutant}
{%
\resizebox{\columnwidth}{!}{
\setlength{\tabcolsep}{2pt}
\begin{tabular}{clccc}
\toprule
\textbf{Design} & \textbf{Mutant description} &
\textbf{AR} & \textbf{DR} & \textbf{DA} \\
\midrule
\multirow{10}{*}{SD}
  & M1: Confidence threshold 3$\to$2      & 1.000 & 0.027 & 1.000 \\
  & M2: Conf.\ decrement off-by-one       & 1.000 & 0.950 & 1.000 \\
  & M3: State FIRST/SECOND swapped        & 1.000 & 0.073 & 1.000 \\
  & M4: Overflow check omitted (s1)       & 0.500 & 0.003 & 0.500 \\
  & M5: \texttt{stride\_2\_valid} inverted & 1.000 & 0.907 & 1.000 \\
  & M6: \texttt{stride\_1\_o} wrong fallback & 1.000 & 0.023 & 1.000 \\
  & M7: \texttt{stride\_2} wrong slot     & 0.000 & 0.000 & 0.000 \\
  & M8: \texttt{stride\_1} never decrements & 1.000 & 0.040 & 1.000 \\
  & M9: \texttt{stride\_2} threshold 3$\to$2 & 1.000 & 0.010 & 1.000 \\
  & M10: Overflow omitted (s2)            & 0.500 & 0.003 & 0.500 \\
\midrule
\multirow{10}{*}{ID}
  & M1: ADD/SUB detection swapped         & 1.000 & 0.238 & 1.000 \\
  & M2: SRL/SRA detection swapped         & 1.000 & 0.135 & 1.000 \\
  & M3: SLT/SLTU ops swapped              & 0.000 & 0.000 & 0.000 \\
  & M4: \texttt{data\_we\_o} not set for stores & 1.000 & 0.098 & 1.000 \\
  & M5: rs1/rs2 register addrs swapped    & 1.000 & 0.548 & 1.000 \\
  & M6: \texttt{illegal\_insn\_o} always~0  & 1.000 & 0.070 & 1.000 \\
  & M7: Store data\_type byte/word swap   & 1.000 & 0.063 & 1.000 \\
  & M8: \texttt{rf\_we\_o} always~0       & 1.000 & 0.788 & 1.000 \\
  & M9: Load data\_type byte/word swap    & 0.000 & 0.000 & 0.000 \\
  & M10: XOR/OR ops swapped               & 0.100 & 0.002 & 0.100 \\
\bottomrule
\end{tabular}
}}
\end{table}

\subsection{LLM4DV Per-Mutant Breakdown}\label{sec:llm4dv_per_mutant_appendix}
Returning to the bug-detection metrics of LLM4DV Section, this Section provides the full per-mutant breakdown (AR, DR, DA for each of the ten individual mutants M1--M10 in the designs). 
Table 9 shows that AR and DR are decoupled at the per-mutant level. Most mutants reach $\mathrm{AR} \geq 0.9$, yet DR frequently falls below $0.1$. Only SD-M2, SD-M5, and ID-M8 exhibit genuine specificity ($\mathrm{DR} = 0.79$--$0.95$); SD-M7, ID-M3, and ID-M9 are never detected; and the remaining mutants fall into a partial-detection band.


\end{document}